\documentclass[aps,prl,twocolumn,superscriptaddress,groupedaddress]{revtex4}  
\usepackage{graphicx}  
\usepackage{dcolumn}   
\usepackage{bm}        
\usepackage{amssymb}   
\usepackage{textcomp}
\usepackage{mathtools}
\usepackage{color}
\usepackage{pgffor}
\begin{document}


\title{Embodiment of Learning in Electro-Optical Signal Processors}

\author{Michiel Hermans}\thanks{michiel.hermans@ulb.ac.be}
\author{Piotr Antonik}\thanks{piotr.antonik@ulb.ac.be}\affiliation{Laboratoire d'Information Quantique, Universit\'e Libre de Bruxelles, 50 Avenue F. D. Roosevelt, CP 225, B-1050 Brussels, Belgium}
\author{Marc Haelterman}\affiliation{Service OPERA-Photonique, Universit\'e Libre de Bruxelles, 50 Avenue F. D. Roosevelt, CP 194/5, B-1050 Brussels, Belgium}
\author{Serge Massar$^1$}

\begin{abstract}
  Delay-coupled electro-optical systems have received much attention for their dynamical properties and their potential use in signal processing. In particular it has recently been demonstrated, using the artificial intelligence algorithm known as reservoir computing, that photonic implementations of such systems solve complex tasks such as speech recognition. Here we show how the backpropagation algorithm can be physically implemented on the same electro-optical delay-coupled architecture used for computation with only minor changes to the original design. We find that, compared when the backpropagation algorithm is not used, the error rate of the resulting computing device, evaluated on three benchmark tasks, decreases considerably. This demonstrates that electro-optical analog computers can embody a large part of their own training process, allowing them to be applied to new, more difficult tasks. \\ 
  \vspace{.1cm} \\
  Published version : http://dx.doi.org/10.1103/PhysRevLett.117.128301
\end{abstract}

\pacs{}
\maketitle

{\em Introduction.}
Nonlinear dynamical systems, such as neural networks (NN), can be used to perform highly complex computations, e.g. speech or image recognition.
One of the main difficulties when using such systems is to train their internal parameters. The backpropagation (BP) algorithm \cite{Rumelhart1986,Werbos1988} is one of the most important algorithms in this area, and 
is behind the remarkable successes achieved in the field of deep learning in the last decade \cite{LeCun2015}. 
The simple idea behind the BP algorithm is to compute the derivative (or gradient) of a cost function in the parameter space of the system. The gradient is then subtracted from the parameters themselves in order to reduce the cost function.
This process is repeated until the cost function no longer reduces.

Such nonlinear dynamical systems can be implemented in hardware. Here also the training of internal parameters is key and the use of the BP algorithm is highly beneficial in order to improve performance  \cite{Hermans2014a,Hermans2015a}. However implementing the BP algorithm in hardware systems can be difficult because of the need of an accurate model to compute the gradient and because of the resources necessary to run the BP algorithm. 
Remarkably, in certain cases the BP algorithm can be implemented physically on the system it is optimising
\cite{Hermans2015}.
The basic idea behind this advance is to use a slightly modified version of the system for propagating error signals backwards, i.e. for running the BP algorithm.
Such self-learning computing systems could be highly advantageous, as any gain in terms of processing speed or limited power consumption will also apply to the training phase. Furthermore having the same hardware computing the BP algorithm eliminates, to a large extent, the need for an accurate model of the system. This idea may conceivably also have implications for biological neural networks, as these are physical system that -- using mechanisms that are not yet well understood -- can both compute and  carry out their own training process.
Reference \cite{Hermans2015} also reported a proof of concept experiment in which physical BP was tested on a simple task, but left open the question of whether the algorithm, with all the imperfections inherent in an experiment, can provide the same improvement in performance as numerical approaches \cite{Hermans2014a,Hermans2015a}.

References \cite{Hermans2014a,Hermans2015a,Hermans2015} used as computational device 
a delay dynamical system (see \cite{Erneux2009,Flunkert2013}). 
Such systems can be exploited to realise  a  form of analog computer based on the Reservoir Computing (RC) paradigm \cite{Jaeger2001,Verstraeten2007a} in which unoptimised high-dimensional dynamic systems (termed \emph{reservoirs}) are used as signal processors.
The RC approach is  simple, versatile and can be applied to a wide set of problems (see the review  \cite{Lukosevicius2009}) and experimental implementations  \cite{Fernando2003,Caluwaerts2011, Appeltant2011,Brunner2013,Larger2012,Paquot2012,Vandoorne2014,Haynes2015,Vinckier2015}. 
Applying the BP algorithm to delay-coupled signal processors allows one to optimise many more parameters than in traditional RC, yielding  significant improvements in performance as was  shown in simulation in \cite{Hermans2014a},
and subsequently in an experiment \cite{Hermans2015a}  in which BP was applied to a numerical model of the system, and the results of the BP algorithm applied to the physical experimental setup.

Here we implement the  BP algorithm physically on an electro-optic delay dynamical system used as signal processor. Our key innovation is to modify the system used in 
\cite{Larger2012,Paquot2012} by adding a photonic setup capable of implementing both the nonlinearity and its derivative, so that it can be used both as signal processor and to perform the BP algorithm.  We test our system on several tasks considered hard in the machine learning community, including a real world phoneme recognition task (the TIMIT task, discussed later in this paper), obtaining state of the art results when the BP algorithm is used. 
The present work thus demonstrates the full potential of physical BP. It  constitutes an important step towards self-learning hardware, with potential applications towards ultra-fast, low energy consumption, computing systems.

In the following we first recall the principles of reservoir computing and error back propagation, before introducing our experimental implementation. We then report the results obtained on several benchmark tasks, and conclude with a discussion of the results and their implications.

{\em Reservoir Computing.}
In typical RC tasks, the goal is to map an input sequence $s_i$ (where $i \in \{1,\cdots,L\}$, with $L$ the total sequence length) to an output sequence $y_i$, which has target values $y^*_i$, for example a speech signal to a sequence of labels. 
In order to use delay-coupled systems as reservoir computers,
the discrete time input sequence $s_i$ is encoded into a continuous time function $z(t)$ by the input mask $m(r)$ and bias mask $m_b(r)$, where $r\in\left[0, T\right]$, with $T$ the \emph{masking period}, as follows
\begin{equation}
z(t)=z(iT+r)=m(r) s_i + m_b(r)
\ .\label{Input:forward}
\end{equation}
In our implementation, we use a delay-coupled system with sine nonlinearity (which stems from the transfer function of the intensity modulator, as will be explained below), which obeys the equation:
\begin{equation}
  a(t + D) =\mu \sin\left(a(t) + z(t)\right) 
  \label{eq:forward}
\end{equation}
where $a(t)$ is the state variable and $D$ is the delay. The factor $\mu$ corresponds to the total loop amplification. Eq. \eqref{eq:forward} can be seen as a special case of the Ikeda delay differential equation \cite{Ikeda1987}.

One then needs to map the continuous time state variable $a(t)$ to a discrete time output sequence $y_i$. This is performed
using an output mask $u(r)$ where $r\in\left[0, T\right]$ and a bias term $u_b$ as follows:
\begin{equation}
y_i  = \int_0^T dr\; a(iT+r)u(r) + u_b
\ .\label{Output:forward}
\end{equation}

In the RC paradigm the input mask is typically chosen randomly, and the output mask $u(r)$ and $u_b$ is determined by solving a linear system of equations which minimises the mean square error $C$ between the desired and actual output: $C = \left<(y_i - y_i^*)^2\right>_i$.

{\em Error Backpropagation.}
The goal of applying error backpropagation to the above scheme is to optimise both the input and output masks $m(r)$, $m_b(r)$, $u(r)$ and $u_b$, knowing the output $a(t)$, and the desired output $y_i^*$. To this end one needs the gradient of $C$ with respect to the masks, given by (the proof is given in the Supplementary Material):
\begin{eqnarray}
\bar{e}(iT+r)&=& e_i u(r)\ ,\label{Input:backward}\\
e(t-D) &=& J(t)\left(e(t) + \bar{e}(t)\right)\ ,\label{DS:backward}\\
J(t)&=&\mu \cos\left(a(t)+z(t)\right)\ ,\label{DSJ(t):backward}\\
\frac{dC}{dm(r)}&=&\sum_i e(iT+r) s_i\ ,\label{Outm:backward}\\
\frac{dC}{dm_b(r)}&=&\sum_i e(iT+r) \ ,\label{Outmb:backward}
\end{eqnarray}
where $\bar{e}(t) = {\partial C}/{\partial a(t)}$ is a continuous time signal and, as above, $ i \in \{1,\cdots,L\}$ and $ r\in\left[0, T\right]$.  
One can then iteratively improve the masks so as to lower $C$.

\emph{Physical BP.}
In order to use the same hardware for both the signal processing and its own training, one exploits the very close analogy between Eqs. (\ref{Input:forward}) and (\ref{Input:backward}) -- both are formed in the same way from a discrete time sequence, multiplied by a periodic mask -- as well as the very close analogy between Eqs. (\ref{eq:forward}) and (\ref{DS:backward}) -- both are delay systems.  
However the equation for $e(t)$ depends on future values, so it needs to be solved backwards in time. In practice one time-inverts $\bar{e}(t)$ and $J(t)$ before computing $e(t)$ to obtain a  linear delayed equation: 
\begin{equation}
e(q+D) = J(q)\left(e(q) + \bar{e}(q)\right)\ .\label{DS:backward2}
\end{equation}
where we use $q$ instead of $t$ to remind oneself that we are dealing with time-inverted signals. 
We also note that $J(t)$, the derivative of the nonlinear function, is a cosine, which can also be implemented using the intensity modulator. Although this property of the sine function is key for this experiment, other types of nonlinearity can be implemented in analogue hardware (see the discussions).

\begin{figure}[t]
\begin{center}
\includegraphics[width=0.4\textwidth]{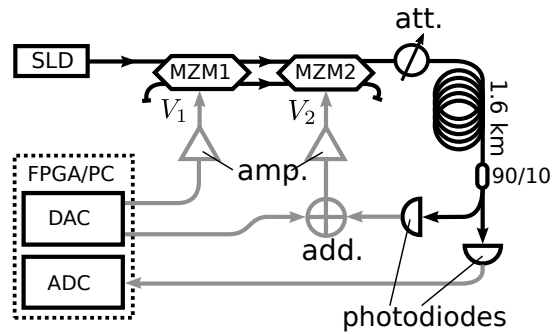}
\caption{Schematic representation of the experimental system. SLD: superluminescent diode; 
  MZM1 and MZM2: dual input - dual output Mach-Zehnder Modulators; $V_1$ and $V_2$: driving voltages of the MZMs;
att.: programmable optical attenuator; add.: electrical combiner; amp.: pulse amplifier.}
\label{fig:physical_setup}
\end{center}
\end{figure}

{\em Experimental implementation.}
In the present work we show how Eqs. (\ref{eq:forward}) and (\ref{DS:backward2}) can be realised using the same physical setup. Our fibre optics experiment is depicted in Figure \ref{fig:physical_setup}. Light is generated by a superluminescent diode (SLD) emitting  in the telecommunications band (1550 nm, with a 33 nm FWHM), modulated by two dual input / dual output Mach-Zehnder modulators (MZM), and attenuated using a programmable optical attenuator used to control the total loop amplification of the system, i.e. $\mu$ in Eq. (\ref{eq:forward}). It then propagates through  an approximately 1.6 km long spool of optical fibre which provides a total loop delay of 7.93 {\textmu}s. The light is split and enters two photodiodes, one of which provides the feedback signal. The signals are produced and recorded by Digital to Analog Converters (DAC) and Analog to Digital Converters (ADC), controlled by a Xilinx Virtex 6  FPGA chip. The FPGA simultaneously generates the input voltage signals and records the output signals. The FPGA communicates with a PC that controls the whole experiment. (Further details on the experimental setup are given in the Supplementary material).

The key innovation with respect to the earlier experiments \cite{Larger2012,Paquot2012} is the use of two dual input / dual output MZMs, 
see Fig. \ref{fig:physical_setup},
which allows to implement both Eqs. (\ref{eq:forward}) and (\ref{DS:backward2}) using the same physical system. 
Taking into account the incoherence of light in the two branches between the modulators (see Supplementary Material for details), the output of the upper branch of MZM2 (see Fig. \ref{fig:physical_setup}) can be found to be:
\begin{equation}
I_2^+ = \frac{I_0}{2}\left[1 +  \sin(V_1/V_0)\sin(V_2/V_0)\right],
\end{equation}
where $I_0$ is the input intensity in the upper branch of MZM1, $V_1$ and $V_2$ are the driving voltages and $V_0$ a constant depending on the MZM. The computational details are presented in the Supplementary Material. 
  In the forward mode, we choose $V_1/V_0 = \pi/2$. The transfer function thus acts as a sinusoidal function for the input argument $V_2/V_0= a(t)+z(t)$. The constant offset $I_0/2$ is removed by the high-pass filter of the amplifier, that drives the MZM. Therefore, once the loop is closed, we end up with Eq. (\ref{eq:forward}).
In the backward mode we drive MZM1 with a voltage $V_1/V_0=a(q) + z(q)+\pi/2$, and MZM2 with a signal proportional to $\bar{e}(q) + e(q)$, but scaled down sufficiently such that $ \sin(V_2/V_0)\approx V_2/V_0 = \bar{e}(q) + e(q)$, which gives the desired functionality for the adjoint system Eq. (\ref{DS:backward2}).

In order to train our reservoir computer, we first choose a value of $\mu$ close to the threshold for instability.  We then iterate the following three steps for (typically) several thousands of iterations, during which performance slowly improves until it converges:

1) We take the training data (typically a small subsequence of the complete set), and convert it to $z(t)$ using the input masks. We feed this signal to the experimental setup, physically implementing Equation \ref{eq:forward}. Next, we measure and record the signal $a(t)$, and generate an output sequence $y_i$ using the output masks.

2) From the output and the desired target values we compute the sequence $e_i=\partial C/\partial y_i$ at the output, and convert it to $\bar{e}(t)$, now using the output mask as an input mask. Next we time-invert it and feed it back into the experimental setup. Simultaneously we drive the first MZM with the (time-inverted) signal $a(q) + z(q)$ in order to implement the online multiplication with $J(q)$. We record the response signal $e(q)$.

3) From the recorded signals $a(t)$ and $e(t)$ we obtain the gradients for the masking signals, which we  use to update the input and output masks:
\begin{eqnarray}
m(r) & \leftarrow & m(r) - \eta\,{dC}/{dm(r)},\nonumber \\
m_b(r) & \leftarrow & m_b(r) - \eta\,{dC}/{dm_b(r)},\nonumber\\
u(r) & \leftarrow & u(r) - \eta\,{dC}/{du(r)},\nonumber\\
u_b & \leftarrow & u_b - \eta\,{dC}/{du_b},
\end{eqnarray}
where $\eta$ is a (typically small) learning rate. In order to speed up convergence we applied a slightly more advanced variant of these update rules known as Nesterov momentum \cite{Nesterov1983,Sutskever2013} (details are given in the Supplementary material).

{\em Results.}
We experimentally validate the above scheme using the system described in Fig. \ref{fig:physical_setup} by testing it on three time series processing task. We consider first of all the NARMA10 task \cite{Atiya2000}, an academic task often used in the RC community. Here the input sequence $s_i$ consists of a series of 
independent and identically distributed random
numbers drawn uniformly from the interval $\left[0, 0.5\right]$. The desired output sequence is  given by 
\[y^*_i= 0.3y^*_{i-1} + 0.05y^*_{i-1}\sum_{n=1}^{10}{y^*_{i-n}} + 1.5s_{i} s_{i-9}+ 0.1.\]
The second task we will call VARDEL5 (from \emph{variable delay}). Here the input sequence consists of i.i.d. digits drawn from the set  $\{1,2,3,4,5\}$. The desired output is then given by $y^*_i = s_{i - s_i}$, i.e., the goal is to retrieve the input instance delayed with the number of time steps given by the current input. 

As a performance metric for NARMA10 and VARDEL5 we use the \emph{normalised root mean square error} (NRMSE), which is given by
\[
\text{NRMSE} = \sqrt{\frac{\left<(y_i - y^*_i)^2\right>_i}{\left<(y^*_i)^2\right>_i}}.
\]
The NRMSE varies between 0 (perfect match), and 1 (no relation between output and target).

The third task is a frame-wise phoneme labelling task. We use the TIMIT dataset \cite{Garofolo1993}, a speech dataset in which each time step has been labelled with one of 39 phonemes. The input data is high-dimensional (consisting of 39 frequency channels), and the desired output is one of (coincidentally) 39 possible output classes. The goal is to label each frame in a separate test set. Consequently, the performance metric is now the classification error rate, i.e., the fraction of misclassified phonemes in the test set. Note that the masking scheme and BP algorithm is easily extended to multidimensional in -- and output sequences (more details are provided in the Supplementary material).
The TIMIT task has been studied before in the context of RC, which has shown it to be challenging, typically requiring extremely large reservoirs to obtain competitive performance \cite{Triefenbach2010,Triefenbach2014}.

\begin{figure}[t]
\begin{center}
\includegraphics[width=0.45\textwidth]{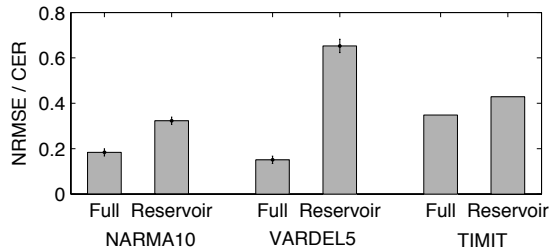}
\begin{picture}(1, 1)
  \put(-235,100){\colorbox{white}{\textcolor{white}{A}}}
\end{picture}
\caption{Comparison of performances for the three tasks under consideration. We show either NRMSE (for NARMA10 and VARDEL5) or the classification error rate (CER) for TIMIT. For each task we show performance for a fully trained systems (Full) vs. those trained using the RC paradigm (Reservoir). Error bars indicate standard deviations if available.} \label{fig:results}
\end{center}
\end{figure}

For all these tasks we compared performance of the fully trained system to traditional RC, where we kept the input and bias masks fixed and random, and only optimised their global scaling and the feedback strength parameter $\mu$. Full experimental details for each task may be found in the supplementary material, together with an example of optimised input masks and of convergence of NRMSE during training. The results are shown in Figure \ref{fig:results}. The experimental setup is successful in performing both useful computations, and implementing its own training process. The fully trained system consistently outperforms the RC approach in all tasks considered.

For the NARMA10 task we improve over all previous experimental results. The previous best was published in \cite{Vinckier2015}, which reported an NRMSE of 0.249 for 50 virtual nodes, and 0.22 for 300 virtual nodes, whereas here we obtain a NRMSE of 0.185 for 80 nodes (note that in \cite{Vinckier2015} they report \emph{normalised mean square error} (NMSE), which is the square of the NRMSE). That result was obtained on an experimental setup that was specially designed to produce a minimal amount of noise (using a passive cavity as a reservoir). The  lowest reported experimental NRMSE on a setup equivalent to ours was 0.41 \cite{Paquot2012}. Note that we obtain a better average performance for the RC setup ($\text{NRMSE}=0.32$), which is most likely due to the higher number of virtual nodes (80 as opposed to 50 in \cite{Paquot2012}).

For the VARDEL5 task, we cannot directly compare to literature, however as pointed out in chapter 5 of \cite{Hermans2012a}, this task is an important example of a task that is so nonlinear that it is nearly impossible to solve it with RC.  This is confirmed here; the NRMSE of RC is 0.66, indicating that the reservoir has only captured the task on a very rudimentary level. The fully trained system shows a drastically better performance (NRMSE = 0.15). This shows that training the input masks not just allows for better performance on existing tasks, but also allows to tackle tasks that are so intricate that they are considered beyond the reach of traditional RC.

For the TIMIT task we obtain a classification error rate of 34.8\% for fully trained systems, vs. 42.9\% for the standard RC approach. These results are only slightly worse than similar experimental results presented in \cite{Hermans2015a}, (33.2\% for fully trained systems and 40.5\% for the RC approach) where 600 virtual nodes were used as opposed as 200 in our case. 

{\em Discussion.}
The present work confirms the results anticipated in \cite{Hermans2014a, Hermans2015a}: 
the performance of delay-based reservoir computers can be drastically improved by optimising both input and output masks. Furthermore, following the proposal of \cite{Hermans2015}, we showed that the underlying hardware is capable of running a large part of its own optimisation process.
We performed our demonstrations on a fast electro-optical system (whose speed could be readily improved by several orders of magnitude, see, e.g. \cite{Brunner2013}), 
and on tasks considered hard in the RC community. Importantly, our work has revealed that the BP algorithm is robust against various experimental imperfections (see the Supplementary Material for details), as the performance gains we obtained on all three tasks were similar to those predicted by numerical simulations.

Although our experiment relies on the sine nonlinearity and its cosine derivative, other nonlinear functions can also be successfully realised in hardware with their derivatives. For instance, the so-called linear rectifier function, which truncates the input signal below a certain threshold, is a popular activation function in neural architectures \cite{glorot2011deep}. Its derivative is a simple binary function which can be easily implemented using an analogue switch, as in \cite{Hermans2015}. 
In \cite{shi2002generator} it is shown how to implement a sigmoid nonlinearity and its derivative.
And in \cite{Vandoorne2014, Vinckier2015} the nonlinearity is quadratic, and therefore the derivative, which is linear, should also be easy to implement. Furthermore, the BP algorithm is robust against imperfect implementation of the derivative, as shown in section 4.3 of the Supplementary Material, and in the Supplementary Material of \cite{Hermans2015} (Supplementary Note 4). Therefore we expect that physical implementation of the BP algorithm will be possible in a wide variety of physical systems.

The current setup still requires some slow digital processing to perform the masking and to compute gradients from the recorded signals. Performing masking operations in analog hardware, however, is actively being researched \cite{Duport2016}, and these approaches could be used to speed up the present setup. Another limitation is the relatively slow data transfer between the FPGA and the computer. Implementing the full training algorithm on the FPGA would drastically increase the speed of the experiment. FPGA's have already been demonstrated to be useful for controlling and training electro-optical signal processors \cite{Antonik2015online,Antonik2016}.

Nowadays, there's an increased interest in unconventional, neuromorphic computing, as this could allow for energy efficient computing, and may provide a solution to the predicted end of Moore's law \cite{Waldrop2016}. These novel approaches to computing will likely be made with components that exhibit strong element-to-element variability, or whose characteristics evolve slowly with time. Self-learning hardware may be the solution that enables these systems to fulfil their potential. The results in \cite{Hermans2015} and in this paper therefore constitute an important step towards this goal.


\begin{acknowledgments}
The authors acknowledge financial support by Interuniversity Attraction Poles Program (Belgian Science Policy) project Photonics@be IAP P7-35, by the Fonds de la Recherche Scientifique FRS-FNRS and by the Action de Recherche Concert\'ee of the F\'ed\'eration Wallonie-Bruxelles through grant AUWB-2012-12/17-ULB9.
\end{acknowledgments}



\end{document}


\maketitle

\section{Experimental setup}

The experimental setup depicted in Fig. 1 of the main text uses the following components: 
\begin{itemize}
    \item Superluminescent diode (SLED): Thorlabs, model SLD1550P-A40), center  wavelength 1550 nm, FWHM 33 nm. 
    \item Dual input/dual output Mach Zehnder modulators (MZM): EOspace, model number AX-2x2-0MSS-12-PFA-PFA.
    \item Programmable optical attenuator: Agilent, model 81571A.
    \item Photodiodes: Terahertz Technologies, model TIA-525.
    \item Pulse amplifiers: Mini-Circuits, model ZHL-32A+.
    \item FPGA, ADC, and DAC: 4DSP FMC151 daughter card containing a two-channel DAC and ADC, controlled by a Xilinx Virtex 6  FPGA chip.
\end{itemize}

Note that the FPGA simultaneously generates the voltage signal that represents $z(t)$ and records the voltage signal representing $a(t)$. The FPGA also performs a minimal signal processing step by selecting and averaging over the middle samples of each masking step (see Section \ref{Mask} for more details). The remaining processing steps are carried out on a PC.

Sending and receiving data to and from the FPGA is currently the main speed bottleneck of the experiment. Even though a single  training iteration lasts only about 0.6 seconds for the NARMA10 and VARDEL5 task, most of this time is spent on the communication overhead with the PC (buffering). If the entire experiment were to be performed on the FPGA (which is feasible), a single training iteration would take of the order of milliseconds.

\section{Online multiplication using cascaded MZMs}\label{sec:MZM}

\begin{figure}
\begin{center}
\includegraphics[width=0.2\textwidth]{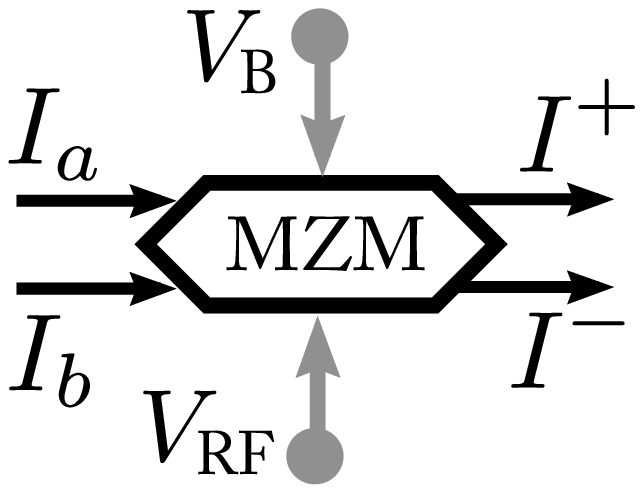}
\caption{Schematic representation of a dual input / dual output Mach Zehnder modulator. The MZM is driven by the sum of two input voltages: one constant bias voltage $V_\textrm{B}$ and a fast signal $V_\textrm{RF}$.} \label{fig:MZM}
\end{center}
\end{figure}

As mentionned in the main text, we use two back to back dual input/dual output Mach-Zehnder modulators for implementation of both Eqs. (2) and (10) from the main text using the same setup.
The main fact we use is that the spectrum of the SLD is narrow enough to allow for a large extinction ratio by the MZMs, but is broad enough that the light in the two branches entering MZM2 from MZM1 can be considered incoherent. In the present experiment, the coherence length of the light from the SLD is of the order of a few hundreds of micrometers, which means that  a very small difference in path length for the connections in between MZM1 and MZM2 is sufficient to make the two signals incoherent.

Consider the operation of a single MZM, schematised in Fig. \ref{fig:MZM}. The intensities of the incoming light sources are denoted by $I_a$ and $I_b$, and the MZM is driven by a voltage $V$, which is the sum of a constant bias voltage  $V_\text{B}$ and a fast voltage signal $V_\text{RF}$. The bias voltage was omitted in the main text to avoid confusion. Taking into account the incoherence between the two input signals, the intensities of the output branches ($I^+$ and $I^-$) are  given by:
\begin{eqnarray}
I^+ &=& I_a\frac{1 + \sin(V/V_0)}{2} + I_b\frac{1 - \sin(V/V_0)}{2},\nonumber\\
I^- &=& I_a\frac{1 - \sin(V/V_0)}{2} + I_b\frac{1 + \sin(V/V_0)}{2},
\end{eqnarray}
with $V_0$ a constant depending on the MZM. 

It is now easy to model the output of the two cascaded MZMs. Suppose the source has an intensity $ I_0$, and no light enters the second input of MZM1. And suppose MZM1 and MZM2 receive voltages $V_1$ and $V_2$, respectively. The output intensities $I_1^+$ and $I_1^-$ of MZM1 are given by
\begin{eqnarray}
I_1^+ &=& I_0\frac{1 + \sin(V_1/V_0)}{2},\nonumber\\
I_1^- &=& I_0\frac{1 - \sin(V_1/V_0)}{2} .
\end{eqnarray}
The intensity $I_2^+$ at the first output branch of MZM2 is then:
\begin{eqnarray}
I_2^+  &=& I_0\frac{(1 + \sin(V_1/V_0))(1 + \sin(V_2/V_0))}{4}  \\
& &+  I_0\frac{(1 - \sin(V_1/V_0))(1 - \sin(V_2/V_0))}{4}\\
&=&\frac{I_0}{2}\left[1 +  \sin(V_1/V_0)\sin(V_2/V_0)\right].
\end{eqnarray}

In the experiment MZM1 receives a constant bias signal on top of an RF driving signal, such that $V_1/V_0 = \pi/2 + V'_1/V_0$, with $V'_1$ the RF signal. We can thus write:
\[I_2^+ = \frac{I_0}{2}\left[1 +  \cos(V'_1/V_0)\sin(V_2/V_0)\right].\]

We use the setup in two modes. In the forward mode, $V'_1 = 0$, so that the cascaded MZMs behave as:
\[I_2^+ = \frac{I_0}{2}\left[1 +  \sin(V_2/V_0)\right],\]
i.e., the transfer function acts as a sinusoidal function for the input argument $V_2/V_0$, which is equal to the sum of the input signal $z(t)$ and the system state $a(t)$. Note that a constant offset $I_0/2$ is added to the output. We use, however, amplifiers with a high-pass filter to drive the MZMs, which remove the DC offset. Therefore, once the loop is closed, this constant bias is removed, and we effectively end up with Eq. (2) in the main text.

In the backwards mode, we drive MZM1 with a voltage $V'_1$ proportional to $a(q-D) + z(q-D)$. MZM2 is driven with a signal proportional to $\bar{e}(q) + e(q)$, but scaled down sufficiently such that $ \sin(V_2/V_0)\approx V_2/V_0 = \bar{e}(q) + e(q)$. This means that in the backwards mode we can write:
\[I_2^+ = \frac{I_0}{2}\left[1 +  \cos\left(a(q + D) + z(q + D)\right)(\bar{e}(q) + e(q)\right],\]
which is (up to the constant bias, and the factor $\mu$ which is imposed later by the optical attenuator) the desired functionality for the adjoint system (see Eq. (10) in the main text).

\section{Derivation of gradients and adjoint system}

\subsection{Setting up the problem}

\begin{figure}[t]
\begin{center}
\includegraphics[width=0.42\textwidth]{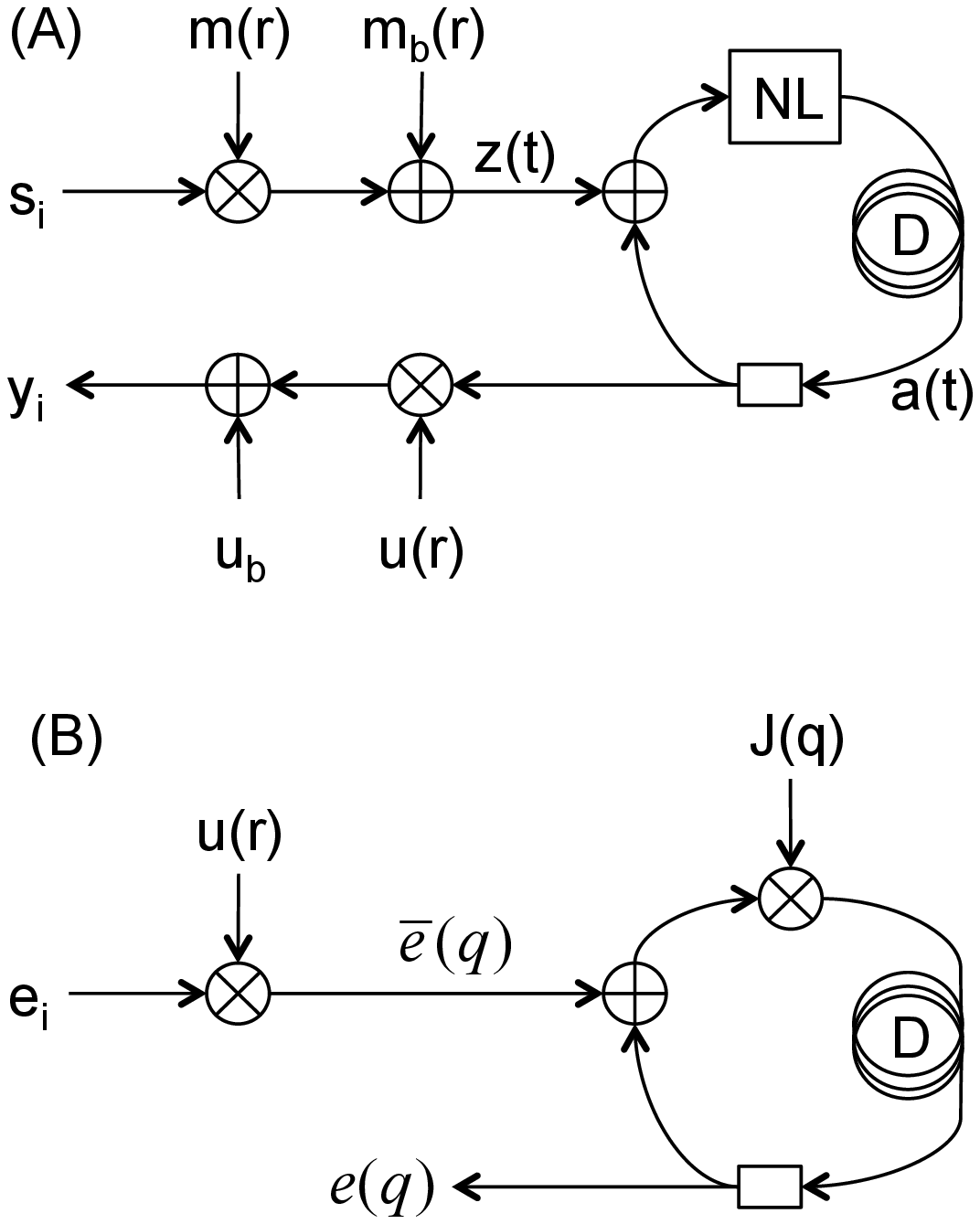}
\caption{\textbf{A}:  Schematic depiction of the forward system, as given in the main text and in Supplementary Material Eqs. (\ref{eq:z(t)}, \ref{eq:at}, \ref{eq:yi}).  \textbf{B}: Schematic depiction of the backward system, as given in the main text and in Supplementary Material Eqs. (\ref{bare}, \ref{eRec}), where $q$ is the backwards time.} \label{fig:concA}
\end{center}
\end{figure}

\begin{figure}[t]
\begin{center}
\includegraphics[width=0.42\textwidth]{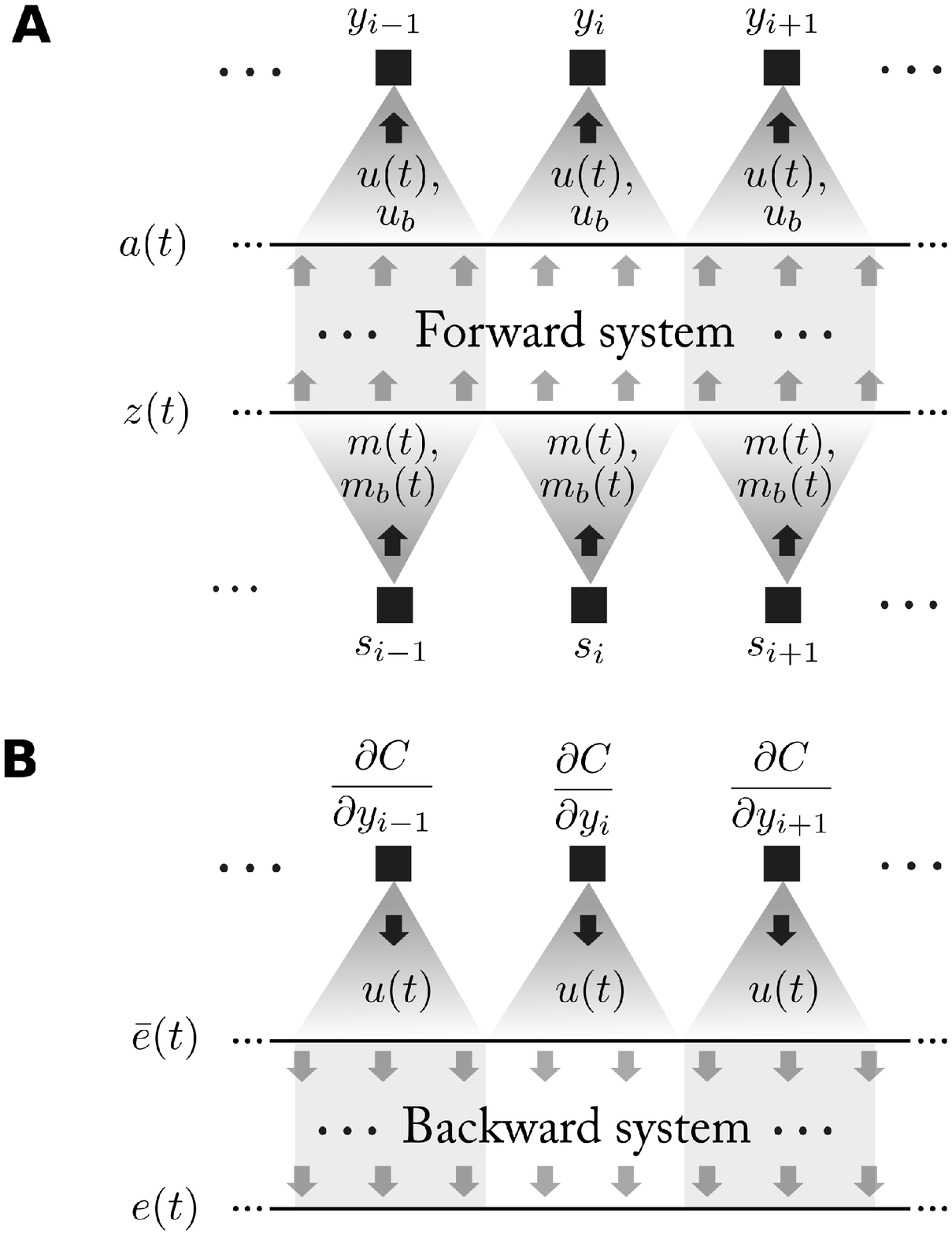}
\caption 
{\textbf{A}:  Schematic depiction of information flow when the system is used in the forward direction. On the bottom, the input sequence $s_i$ is converted to a continuous-time signal $z(t)$ (with time running from left to right). Each instance in the sequence is multiplied with the finite-length masking signal $m(t)$ and added to $m_b(t)$. These sequences are then concatenated in time to form $z(t)$, the input to the forward system. The output $a(t)$ of the forward system is then converted into an output sequence $y_i$ by segmenting $a(t)$ in time, and multiplying the segments with the output masks $u(t)$, and integrating over each of them. \textbf{B}: Schematic depiction of the information flow in the ``backwards'' mode. The derivatives ${\partial C}/{\partial y_i}$ are used as an input sequence. They are multiplied by $u(t)$ which now plays the role of input mask. This yields the  signal $\bar{e}(t)$ that serves as input for the backward system. The output of the backward system is $e(t)$.} \label{fig:concepts}
\end{center}
\end{figure}

We wish to find the gradient of a cost function $C$ w.r.t. the parameters that can be optimised. In order to achieve this we have to use the chain rule through all the dependencies that describe the system. We will then obtain the backward equations  given in the main text. Figure \ref{fig:concA} gives a schematic of how the forward and backward equations must be implemented experimentally. Figure \ref{fig:concepts} depicts the information flow in the forward and backward systems.

We first recall the relevant equations describing the forward system.
The input signal $z(t)$, formed by concatenating the input masks weighted with the current input sample $s_i$ can be rewritten as
\begin{equation} z(t) = s_{\lceil t/T \rceil}m(t\text{ mod }T) + m_b(t\text{ mod }T),\label{eq:z(t)}
\end{equation}
where $\lceil . \rceil$ indicates the ceiling function, so that $\lceil t/T \rceil=i$ gives the index of $s_i$ corresponding to the time $t$. We use the modulo operation in the argument of the input masks to indicate that the masks are repeated over time. Next we write down the expression for the reservoir state $a(t)$:
\begin{equation}
a(t + D) = \mu\sin\left(a(t) + z(t)\right).\label{eq:at}
\end{equation}
Finally, we can write the formula for the output instances $y_i$ as follows:
\begin{equation}
y_i = u_b +  \int_0^T dr\;u(r)a_i(r),\label{eq:yi}
\end{equation}
with $a_i(r)$ = $a(r + (i-1)T)$, the $i$-th segment of the recording of $a(t)$.

In what follows, for the sake of generality and of simplicity of notation, we take the input and output masks to be continuous functions of time. We denote functional derivatives with respect to time dependent functions as ordinary derivatives. The case, relevant to practical implementations, in which the masks depend on a finite number of parameters, is discussed in section \ref{Mask}.
For simplicity in the derivations we will assume, unless indicated otherwise, that all variables, both in continuous time $t$ and discrete time $i$, are defined for $i$ and $t$ going from $-\infty$ to $\infty$. If we have a specific finite input sequence $s_i$ with $i\in\{1,\cdots,L\}$, we simply extend this beyond these bounds assuming that all extra $s_i$ are equal to zero. Similarly, we assume that $z(t)$ is zero if $\lceil t/T \rceil\notin\{1,\cdots,L\}$. Subsequently, if we sum or integrate over $i$ or $t$ without indicating limits, this indicates a summation or integration from $-\infty$ to $\infty$. In practice it turns out that if we only have a finite sequence, we only need to compute states over its corresponding time span. Similarly, when performing backpropagation, we only need to compute backwards over the same time span. All states outside of this interval do not influence the gradient computation, which means there are no problems in considering only finite intervals. This matters as in realistic training scenarios we typically train on relatively short sequences (in the case of the present paper of length 100).
\subsection{Output mask gradient}
For the output masks we can write 
\begin{equation}
\frac{dC}{du(r)} = \sum_i\frac{\partial C}{\partial y_i}\frac{dy_i}{du(r)}.\label{eqdCdu}
\end{equation}
For example, if the cost function we wish to minimise is the squared error over the interval of the input sequence:
\[C = \sum_{i=1}^{L} (y_i - y_i^*)^2,\]
\[
e_i=\frac{\partial C}{\partial y_i} = 2(y_i - y_i^*)\textrm{     for     }i\in\{1,\cdots,L\}.
\]
\[
\frac{\partial C}{\partial y_i} = 0\textrm{     for     }i\notin\{1,\cdots,L\}.
\]
The second factor in Eq. (\ref{eqdCdu}) we can get from Equation \ref{eq:yi}:
\[
\frac{dy_i}{du(r)} = a_i(r),
\]
such that the gradient for the output mask $u(t)$ is simply given by 
\[
\frac{dC}{du(r)} = \sum_i\frac{\partial C}{\partial y_i}a_i(r),
\]
or, given the fact that ${\partial C}/{\partial y_i} = 0$ outside the interval in which the sequence is defined:
\begin{equation}
\frac{dC}{du(r)} = \sum_{i=1}^{L}\frac{\partial C}{\partial y_i}a_i(r).\label{eq:gradut}
\end{equation}
Similarly we find that
\[\frac{dC}{du_b} = \sum_{i=1}^{L} \frac{\partial C}{\partial y_i}.\]

\subsection{Input mask gradient}
The case of the input masks is  more involved. Working out the chain rule we find:
\begin{eqnarray}
\frac{dC}{dm(r)} &=& \sum_i{\frac{\partial C}{\partial y_i}}\frac{dy_i}{dm(r)}.\nonumber\\
&=& \sum_i{\frac{\partial C}{\partial y_i}}\int dt'\;\frac{\partial y_i}{\partial a(t')}\frac{d a(t')}{dm(r)}.\nonumber\\
& = & \int dt'\;\bar{e}(t')\frac{d a(t')}{dm(r)}\label{eq:grad1},
\end{eqnarray}
where we have used 
\[
 \bar{e}(t') =  \frac{\partial C}{\partial a(t')}  = \sum_i{\frac{\partial C}{\partial y_i}}\frac{\partial y_i}{\partial a(t')}\ .
\]
From Equation \ref{eq:yi} we can obtain (using a modulo function in the argument of $u(r)$):
\[\frac{\partial y_i}{\partial a(t')} = \delta_{i,\lceil t'/T \rceil}u(t' \text{ mod }T),\]
i.e., equal to zero when $t'$ did not fall in the segment of time used to produce $y_i$, and equal to the output mask otherwise. This yields:
\begin{eqnarray}
\bar{e}(t') &=& u(t'\text{ mod }T)\frac{\partial C}{\partial y_{\lceil t'/T \rceil}} \nonumber\\
&=& u(r) e_i
\label{bare}
\end{eqnarray}
where $r=t'\text{ mod }T$ and $i=\lceil t'/T \rceil$. 
In other words, $\bar{e}(t)$ is produced by masking the sequence $\partial C/\partial y_i$ with the output mask $u(r)$.

The second factor in Equation \ref{eq:grad1} we work out as follows. Using the chain rule we get 
\begin{equation}
\frac{da(t')}{dm(t)} = \int dt''\;\frac{da(t')}{dz(t'')}\frac{dz(t'')}{dm(t)}.\label{eq:inte}
\end{equation}
and
\[
\frac{da(t')}{dz(t'')}=\frac{\partial a(t')}{\partial z(t'')} + \int dt'''\;\frac{\partial a(t')}{\partial a(t''')}\frac{da(t''')}{dz(t'')}.
\]
From equation \ref{eq:at} we obtain the partial derivatives:
\begin{eqnarray*}
\frac{\partial a(t')}{\partial z(t'')}& =& \frac{\partial a(t')}{\partial a(t'')} \\
&=& \mu\delta(t' - t'' - D)\cos(a(t'-D) + z(t'-D)),
\end{eqnarray*}
Or, more compactly: 
\[\frac{\partial a(t')}{\partial z(t'')} = \delta(t' - t'' - D)J(t'),\]
with 
\[J(t') = \mu\cos(a(t'-D) + z(t'-D)).\] 
This yields
\[
\frac{da(t')}{dz(t'')} = J(t')\left[\delta(t'-t''-D) + \frac{da(t'-D)}{dz(t'')}\right].
\]
By filling in the expression for ${da(t'-D)}/{dz(t'')}$ recursively we can write this as:
\begin{equation}
\frac{da(t')}{dz(t'')} =
\sum_{i=0}^\infty \left[\delta(t'-t''-iD)\prod_{j=0}^{i-1}J(t'-jD)\right] .\label{eq:dadz}
\end{equation}
By filling in Equation  \ref{eq:dadz} in Equation \ref{eq:inte}, and inserting the result in Equation \ref{eq:grad1} we obtain:
\begin{equation}
\frac{dC}{dm(r)} = \int dt'\;dt''\;\bar{e}(t')\sum_{i=0}^\infty \delta(t'-t''-iD)\prod_{j=0}^{i-1}J(t'-jD)\frac{dz(t'')}{dm(r)} .\label{eq:grad2}
\end{equation}
We can solve the integral over $t'$ explicitly. We denote this by $e(t'')$:
\begin{eqnarray}
e(t'')&=&\int dt'\;\bar{e}(t')\sum_{i=0}^\infty \delta(t'-t''-iD)\prod_{j=0}^{i-1}J(t'-jD)\nonumber\\
&=&\sum_{i=0}^\infty\bar{e}(t'' + iD)\prod_{j=0}^{i-1}J(t'' + (i-j)D)\nonumber\\
&=&\sum_{i=0}^\infty\bar{e}(t'' + iD)\prod_{j=1}^{i}J(t'' + jD)\label{eq:eexpression}.
\end{eqnarray}
It's straightforward to prove that $e(t)$ is equal to the expression as presented in the main text (with arguments shifted by $D$):
\begin{equation}
e(t) = J(t+D)(e(t+D) + \bar{e}(t+D)).
\label{eRec}
\end{equation}
Indeed, if we recursively fill in the expression for $e(t+D)$ in Eq. \ref{eRec}, we obtain Eq. \ref{eq:eexpression}. Using this we can reduce Equation \ref{eq:grad2} to
\[
\frac{dC}{dm(r)} = \int dt''\;e(t'')\frac{dz(t'')}{dm(r)}.
\]
From the expression of $z(t)$ we find that
\[
\frac{dz(t'')}{dm(r)} = \delta(t''\text{ mod }T - r)s_{\lceil t/T \rceil}.
\]
Inserting this we can find the final expression for the gradient for the input mask:
\[
\frac{dC}{dm(r)} = \sum_i s_ie_i(r),
\]
or, again using the fact that we defined $s_i = 0$ for $i\notin\{1,\cdots,L\}$:
\begin{equation}
\frac{dC}{dm(r)} = \sum_{i=1}^Ls_ie_i(r),\label{eq:gradmt}
\end{equation}
with $e_i(r) = e(r - (i-1)T)$, the $i$-th segment of the time trace of $e(t)$.
Similarly for $m_0(t)$ we can write:
\begin{equation}
\frac{dC}{dm_b(r)} = \sum_{i=1}^L e_i(r).\label{eq:gradm0}
\end{equation}

\subsection{Multiple inputs/outputs}
The above explanation is easily extended to multiple input and output dimensions. Suppose we have a multivariate time series $\mathbf{s}_i$, where the $k$-th element at time step $i$ is denoted by $\mathbf{s}_i[k]$. We can then easily construct $z(t)$ by defining as many input masks $m_k(t)$ as there are input dimensions and adding them all up:
\[
z(t) = \sum_k \mathbf{s}_{\lceil t/T \rceil}[k]m_k(t\text{ mod }T) + m_b(t\text{ mod }T),
\]
The desired output can similarly exist of a multivariate time series with elements $\mathbf{y}^*_i[l]$. To produce an output $\mathbf{y}_i[l]$ we simply define an output mask $u_l(t)$ and bias $u_l^0$ for each output channel:
\[
\mathbf{y}_i[l] = u_l^0 +  \int_0^T\;dt u_l(r)a_i(r).
\]

The same procedure can now be used to determine the gradients with respect to the multivariate input and output masks.
We find:
\begin{equation}
\frac{dC}{du_l(r)} = \sum_{i=1}^L \frac{dC}{d\mathbf{y}_i[l]}a_i(r),\label{eq:grad_outp_md}
\end{equation}
and
\begin{equation}
\frac{dC}{du_l^0} = \sum_{i=1}^L \frac{dC}{d\mathbf{y}_i[l]}.\label{eq:grad_outp_b_md} 
\end{equation}
The source of the BP equation is now
\begin{equation}
\bar{e}(t') = \sum_l u_l(t'\text{ mod }T)\frac{dC}{dy_{\lceil t'/T \rceil}[l]} \ ,
\label{barel}
\end{equation}
the recurrence for the error $e(t)$, eq. (\ref{eRec}),  is unchanged, and one has
\[
\frac{dC}{dm_k(r)} = \sum_{i=1}^L \mathbf{s}_i[k]e_i(r) \ .
\]

\section{Implementation details}

\subsection{Mask parametrisation}\label{Mask}
While the aforementioned theory is generally valid for continuous-time signals, an experimental setup is limited by the finite bandwidth of the DAC/ADC, and the analog electronic parts. To make sure that these effects play a limited role, we parametrise the input and output masks as piecewise constant functions, which has been common practice for reservoirs of this type \cite{Paquot}. To this end we divide the delay $D$ into an integer number $N_D$ of equal time segments, called \emph{masking steps}. Next we ensure that the masking period $T$ has a total duration that is also contains an integer number $N_T$ of masking steps, in our case one less than the delay: $N_T = N_D - 1$. This allows for the mixing of the states over time, as detailed in \cite{Paquot}.

The input and output masks are picked to be constant for the duration of each masking step. This implies that $z(t)$ is piecewise constant. The fact that both $T$ and $D$ are an integer number of masking steps makes that changes in $a(t)$ only occur in between the masking steps, i.e., they are synchronised with the masking steps, and this is valid for the backwards pass too. In short, $a(t)$, $\bar{e}(t)$ and $e(t)$ are all piecewise constant signals, with values that remain constant during each masking step.

In practice this allows us to reduce effects of noise by averaging the signals representing $a(t)$ and $e(t)$ over several measuring samples during a single masking step. Typically we pick a set of samples from the middle of each masking step, and discard those at the beginning and the end as they may contain artefacts caused by the limited bandwidth of the ADC. More importantly, it allows us to make a discrete time approximation of the entire system. For example, let's consider equation \ref{eq:yi}. The mask $u(t)$ is made up of $N_T$ constant segments of equal length, with values during the segments denoted $u_k$. Similarly, each segment $a_i(r) = a\left(t - (i-1)T\right)$ is piecewise constant, with values we can for example denote with $a^i_k$. The integral reduces to 
\[
y_i = u_b + \sum_{k=1}^{N_T} a^i_ku_k,
\]
(where we absorbed the factor $T$ that emerges from the integration into the values $u_k$). Each particular value $a^i_k$ can be interpreted as the state of the $k$-th `neuron' or `node' state during the $i$-th instance of the input sequence. We can still use the expressions for the gradients in Equations \ref{eq:gradut}, \ref{eq:gradmt} and \ref{eq:gradm0}. Indeed, by construction, the gradient for the output mask $u(t)$ for the duration of a single masking step is a constant (as $a(t)$ remains constant over the segment). The same holds for the gradients for the input masks. This implies that $u(t)$  and $m(t)$ remain piecewise constant during training, and we can in practice describe them simply as lists of values instead of a continuous-time function.

Note that the choice of dealing with bandwidth limitations by using piecewise constant functions is not the only possible avenue. One alternative would be to impose bandwidth constraints on $m(t)$ and $u(t)$, such that the finite signal generator bandwidth and sampling rates form no obstacle in treating the setup as a continuous-time setup. We chose the piecewise-constant constraint as it is more directly related to existing implementations of delay-coupled electro-optical signal processors, and it allows to identify a specific number of `virtual nodes' (the number of segments within the masking period $T$). In other words, the choice of $N_T$ determines the `complexity', or the number of degrees of freedom of the system.

\subsection{Gradient descent}
We used stochastic gradient descent to train the masks; each iteration we drew a 100 time step sequence to determine a gradient. This sequence was either generated on the fly (in the case of VARDEL5 and NARMA10), or drawn randomly from a training set (TIMIT). Note that as the BP equation is linear, we are in principle free to rescale $\bar e$ as we wish. In practice, in order to keep MZM2 in the linear regime, we scaled the input error signal $\bar{e}(t)$ by dividing it by its standard deviation and multiplying with a factor 0.1. 
The learning rate $\eta$ we choose equal to 0.25 at the start of the training process, after which it drops linearly to zero throughout the course of the experiment. On top of that we use Nesterov momentum with a momentum factor 0.9 to speed up convergence \cite{Nesterov,Sutskever}. Nesterov momentum is a heuristic method that finds widespread use in speeding up convergence of stochastic gradient descent. The idea of momentum in gradient descent is to give parameter updates a certain inertia, meaning that previous parameter updates still count in the current one, which helps with overcoming local minima and speeds up convergence. Nesterov momentum is a simple variation of this principle, where the algorithm measures the gradient one update step ahead in order to change its momentum ``ahead of time''.

\subsection{Robustness}
Our work shows that physical BP is robust against imperfections of the physical setup, as illustrated by the following imperfections we were confronted with. 

The first imperfection was the high-pass filtering operation of the amplifiers used to drive the MZMs, with a cut-off frequency of 20 kHz. While the high-pass filter is a desirable property (to get rid of voltage bias), this corresponds to a typical time scale of about 8 \textmu s, which is about the same as the loop delay and therefore not negligible. The current experimental setup does not take this filtering operation into account explicitly. 

A second imperfection was an imbalance in losses between the two fibres connecting MZM1 with MZM2. A third imperfection was that the system was not perfectly linear during the backwards pass, since MZM2 is never a perfectly linear system. There's also an important trade-off here. One can reduce the residual nonlinearity by reducing the amplitude of the incoming voltage signal that represents $\bar{e}(t)$. But in turn this also reduces the signal-to-noise ratio of the measurement during the backpropagation phase, such that one needs to find a good balance between these two effects.

All these effects are imperfections inherent to the physically implemented backpropagation phase, but both in simulation and in the actual experiments we found that they only had a very minor impact on the training process and the overall performance. 

One parameter that turned out to be crucial was the bias voltage of MZM2. The reason is that even a small offset from an effectively zero level introduces a systematic error in the backpropagation process, such that the measured signal (denoted as $e_c(t)$ to indicate that it is corrupted) becomes :
\[
e_c(t-D) = J(t)(e_c(t) + \bar{e}(t) + \widetilde{e}),
\]
with $\widetilde{e}$ a constant offset caused by an incorrectly set voltage bias of MZM2. It turned out that, in the experiments, keeping this bias level effectively equal to zero was  difficult; very slight drifts on the effective working point of the MZM occurred over the course of minutes/hours. Luckily, the backpropagation is a  linear process. This means that we can recover $e(t)$ by performing a second measurement right after measuring $e_c(t)$:
\[
e_r(t - D) = J(t)(e_r(t) + \widetilde{e}),
\]
and 
\[
e(t) = e_c(t) - e_r(t).
\]
In other words we simply need to perform two measurements after each other, where in the second one we send a `zero' input error, and subtract this from the first measurement in order to remove the influence of the offset of MZM2. This turned out to solve the problem.

\section{Tasks}

\begin{figure}
  \centering
  \subfigure[Input mask $m$]{\includegraphics[width=0.45\textwidth]{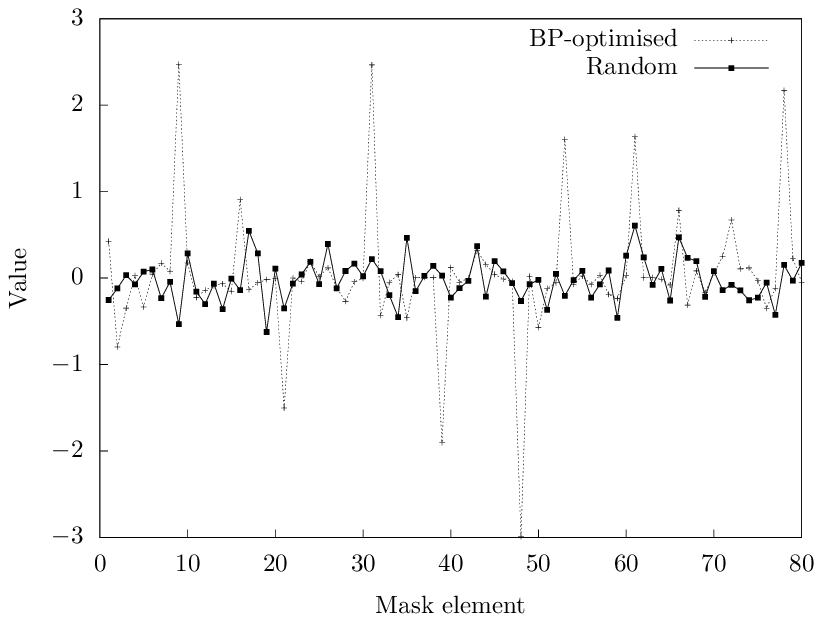}}
  \subfigure[Bias input mask $m_b$]{\includegraphics[width=0.45\textwidth]{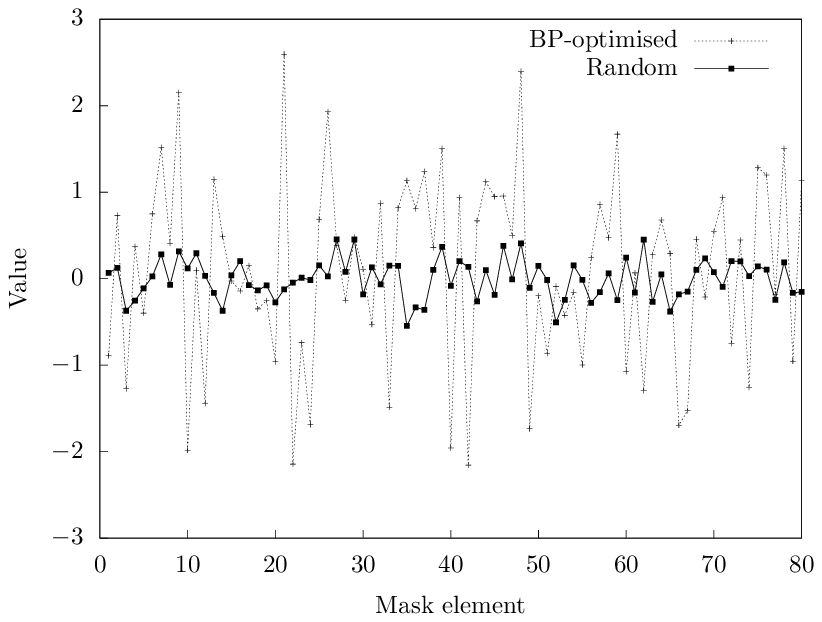}}
  \caption{
    Comparison of BP-optimised input masks (dotted curves) and random RC masks (solid curves) for the VARDEL5 task, for $m(r)$ (top panel) and $m_b(r)$ (bottom panel).
  }
\label{FigMasks}
\end{figure}

\subsection{NARMA10 and VARDEL5}
In the case of NARMA10 and VARDEL5 we divided $T$ into 80 equal time intervals ($N_T=80$), which allowed us to take 16 samples during each masking step, where we averaged over the middle 8 in order to get piecewise-constant values for $a(t)$ and $e(t)$. 
We chose the number of training iterations at 10,000, 20,000 for VARDEL5 and NARMA10, respectively, chosen heuristically as a trade-off between the time required for an experiment and the final performance. (A single iteration lasted approximately 0,6 s).

The cost functions used for NARMA10 and VARDEL5 are the aforementioned sums of squared errors. We repeated the training cycles 10 times, each time with different random input mask initialisations. Output masks were always initialised at zero. For all backpropagation experiments we set the feedback parameter strength parameter $\mu$ effectively equal to one (such that the system is at the `edge of stability'), which we found to give the best performance.

\begin{figure}
  \centering
  \includegraphics[width=0.45\textwidth]{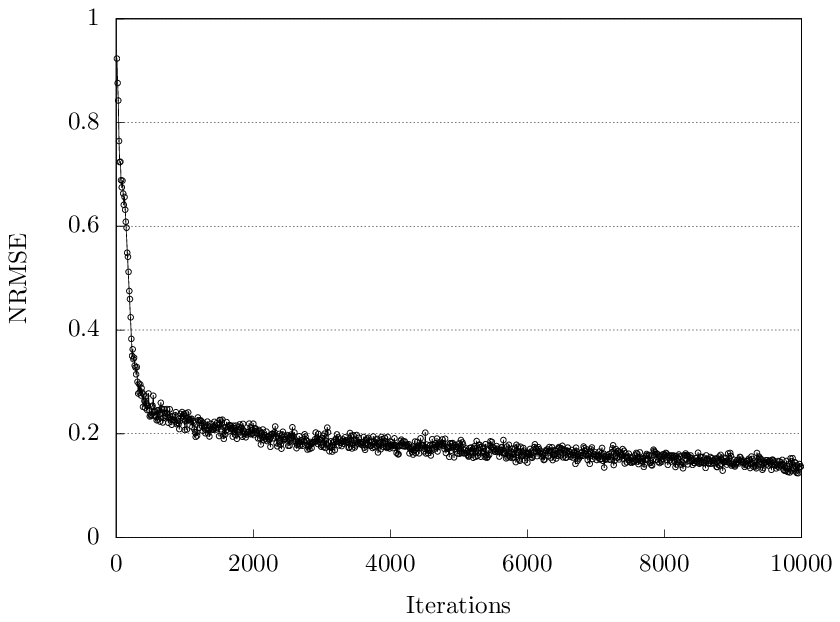}
  \caption{Evolution of the NRMSE during the training process on the VARDEL task. The error falls sharply during the first 300 iterations, and then converges slowly towards $0.15$.}
\label{FigConvergence}
\end{figure}

In Figure \ref{FigMasks} we depict the masks $m(r)$ and $m_b(r)$ for the RC implementation (when they are chosen at random), and after optimisation using the BP algorithm, for the VARDEL5 task. One sees that the BP algorithm dramatically changes the input masks. In particular the mask $m$ is very large at some specific values of $r$, and almost zero for other values. This suggests that in some sense what the optimised reservoir is doing  is storing the value of the input on specific neurons, and then keeping it in memory for some time, before mixing it nonlinearly  with the input several time steps in the future.
In Figure \ref{FigConvergence} we depict how the NRMSE converges over time for the VARDEL5 task, as the BP algorithm slowly improves the input and output masks.

For the reservoir computing results we measured average performance as a function of three scaling parameters: the feedback strength parameter $\mu$ and the scaling of the input mask and bias mask. Once optimal parameters were determined we ensured that the output masks were trained on an unlimited amount of input training data (in practice we observed the test error for increasing amounts of training data, and stopped as soon as the performance no longer improved). This was to ensure that we have a fair comparison to the backpropagation setup, where we generate unlimited amounts of data too. Each experiment was is repeated 10 times, giving rise to the error bars in Figure 3A in the main text.

\subsection{TIMIT}
For the TIMIT task we used 1,000,000 training iterations. Because of this large number of iterations we only performed a single full training cycle.

Measurement noise plays a smaller role in a classification task such as this one, and we divided $T$ into 200 masking steps, taking 8 samples in each and averaging over the middle 4, thereby increasing the number of virtual nodes $N_T$ while taking into account the hardware constraints (sample rates of the DAC and ADC).
Most likely this number can be increased further for example using only 4 samples per masking steps and averaging over the middle 2. In practice we are also limited by the relatively slow communication between the PC and the FPGA, and increasing $N_T$ increases the amount of data that needs to be transferred, slowing down the experiment considerably. Currently, for 1,000,000 iterations the training took two weeks to complete.

We picked $\mu$ at a value slightly under one, but we found in simulations that performance did not strongly depend on it for a broad range of values.

In the case of TIMIT, the goal is to minimise a classification error rate, which is not directly differentiable. One possible strategy is simply to try and minimise the MSE between the output and the target labels (1 for the correct class, zero for all others). Classification would then be performed by the \emph{winner-take-all} approach, where we simply select the output channel with the highest output as the `winner'. In practice, using MSE for classification suffers from some drawbacks. Most importantly MSE will put a lot of emphasis on producing the exact target values (close to zero or one), while we are only interested in performance after selecting the highest output. A better approach is to use a softmax function at the output, which converts the output values into a set of probabilities, and minimise the cross-entropy with the target probabilities (again, 1 for the correct class and zero for all others) . Details on this strategy can be found for example in \cite{Bischop}. In practice the conversion of the output $\mathbf{y}_i[k]$ into probabilities is performed using the so-called \emph{softmax} function:
\[
\mathbf{p}_i[k] = \frac{\exp(\mathbf{y}_i[k])}{\sum_l\exp(\mathbf{y}_i[l])},
\]
Tthe cost function is the cross-entropy: 
\[
C = -\sum_{i=1}^L\sum_k\mathbf{t}_i[k] \ln\mathbf{p}_i[k],
\]
where we denote the target outputs as $\mathbf{t}_i[k] $.
It can then be shown that
\[
\frac{dC}{d\mathbf{y}_i[k]} = \mathbf{p}_i[k] - \mathbf{t}_i[k],
\]
(and again zero if $i\notin\{1,\cdots,L\}$)
This means that the error we have at the output takes on virtually the same form as before, only this time there is the intermediary step of the softmax function. Gradients for the output masks are almost the same as before, except for equations \ref{eq:grad_outp_md} and \ref{eq:grad_outp_b_md} where we use $\mathbf{p}_i[k] - \mathbf{t}_i[k]$ instead of $\mathbf{y}_i[k] - \mathbf{y}^*_i[k]$. As far as the rest of the BP algorithm goes, we now simply have to mask these `output errors' to produce $\bar{e}(q)$, and the rest plays out exactly as before.

For the RC approach, optimising the parameters (input scaling, bias scaling and feedback gain) on the hardware would be too costly in terms of time. Therefore we optimised them on a PC using a simulation of the physical setup. Once we decided on the parameters, we ran all the TIMIT data through the physical setup and recorded all the responses. Next we trained output weights, again using gradient descent with the above cross-entropy loss.